\documentclass{article}
\usepackage{amsmath}
\usepackage{amsthm}
\usepackage{hyperref}
\usepackage[noabbrev]{cleveref}

\newtheoremstyle{mddefinition}
  {-.25\topsep}
  {-.25\topsep}
  {\normalfont}
  {}
  {\bfseries}
  {.}
  {.5em}
  {}

\usepackage[framemethod=Tikz]{mdframed}
\mdfsetup{skipbelow=0pt}
\mdfdefinestyle{defstyle}{
    outerlinewidth = 1,
    roundcorner = 2,
    leftmargin = 7,
    rightmargin = 7,
    backgroundcolor = green!4
}
\mdfdefinestyle{thmstyle}{
    outerlinewidth = 1,
    roundcorner = 8,
    leftmargin = 7,
    rightmargin = 7,
    backgroundcolor = cyan!5
}

\theoremstyle{definition}
\newtheorem{problem}{Problem}
\theoremstyle{mddefinition}
\newmdtheoremenv[nobreak=true, outerlinewidth=0.7]{problem*}[problem]{Problem}

\newenvironment{solution}[1][]
  {%
  \begin{oldproof}[\ifx&#1&Solution\else#1\fi]}
  {\end{oldproof}}

\renewenvironment{proof}
  {\begin{oldproof}}
  {\end{oldproof}}

\crefname{problem}{problem}{problems}
\crefname{problem*}{problem}{problems}

\hypersetup{colorlinks,
    linkcolor={blue},
    citecolor={blue!50!black},
    urlcolor={blue!80!black}}

\usepackage[shortlabels,inline]{enumitem}
\usepackage{amsmath}
\usepackage{amssymb}

\let\abs\relax

\usepackage{mathtools} 
\DeclarePairedDelimiter{\set}{\{}{\}}
\DeclarePairedDelimiter{\abs}{\lvert}{\rvert}

\newcommand\R{\ensuremath{\mathbb{R}}}

\newcommand\mcT{\ensuremath{\mathcal{T}}}

\newcommand\mcV{\ensuremath{\mathcal{V}}}

\usepackage{newfloat}
\usepackage{algorithmicx}
\usepackage[indLines=false, italicComments=false, beginLComment=---~, endLComment=]{algpseudocodex}

\newcommand\textul[1]{\vphantom{#1}\underline{\smash{#1}\vphantom{\tiny ,}}}
\algdef{SxNi}[fn]{Fn}{EndFn}{1em}[2]{\textul{\textsc{#1}(#2):}}%

\algrenewcommand\algorithmicfor{for}
\algrenewcommand\algorithmicwhile{while}
\algrenewcommand\algorithmicdo{}
\algrenewcommand\algorithmicend{end}
\algrenewcommand\algorithmicif{if}
\algrenewcommand\algorithmicthen{}
\algrenewcommand\algorithmicelse{else}
\algrenewcommand\algorithmicreturn{return}

\usepackage{varwidth}
\newenvironment{algo}[1][0.8]
{\fontfamily{bch}\selectfont\small
\linespread{1.0}
\begin{center}\begin{tabular}{|l|}
    \hline
    \hspace{-0.85em} \begin{varwidth}{#1\textwidth}
        \small\;\\[-1.6em]
        \begin{algorithmic}}
{       \end{algorithmic}\;\\[-2.3em]
    \end{varwidth} \\
    \hline
\end{tabular}\end{center}}

\usepackage{todonotes}
\usepackage{geometry}
\usepackage{tikz}
\usepackage{pgfplots}
\pgfplotsset{compat=1.18}
\usetikzlibrary{arrows,shapes,positioning,shadows,trees,math}
\usetikzlibrary{calc}
\usetikzlibrary{shapes}
\usetikzlibrary{shapes.geometric, arrows}
\usetikzlibrary{positioning}
\usepackage{subcaption}

\DeclareMathOperator{\NN}{\mathcal{NN}}

\title{Report on Nearest Dominating Point Queries}
\author{Naman~Mishra, Sreeramji~K~S}
\date{November 2024}

\begin{document}

\maketitle

\section{Introduction}
Given two points $p, q \in \R^d$, we say that $p$ \emph{dominates} $q$ and write
$p \succ q$ if each coordinate of $p$ is larger than the corresponding coordinate of $q$.
That is, if $p = (p^{(1)}, p^{(2)}, \dots, p^{(d)})$ and $q = (q^{(1)}, q^{(2)}, \dots, q^{(d)})$,
$p \succ q$ if and only if $p^{(i)} > q^{(i)}$ for all $1 \le i \le d$.

For example, $p$ and $q$ could represent various ratings for $2$ restaurants, based
on different metrics like taste, affordability, or perhaps
ratings on different platforms, \textit{et cetera}.
$p \succ q$ then means that the first restaurant outperformed the second on each
metric.
Any restaurant owner would like to know restaurants that dominate it, and
would be interested in the closest such restaurants.
This brings up the idea of a \emph{real space}, where restaurants are located,
and a \emph{feature space}, where the ratings of each restaurant live.

Given a list of restaurants and their rating, we solve the problem of determining,
for each restaurant, the closest restaurant to it that dominates it.
We state the problem more formally in the next section, where we tackle a special
case first and move on to a more general case.
We improve upon the algorithm under some assumptions in section 3.

\section{The Problem}
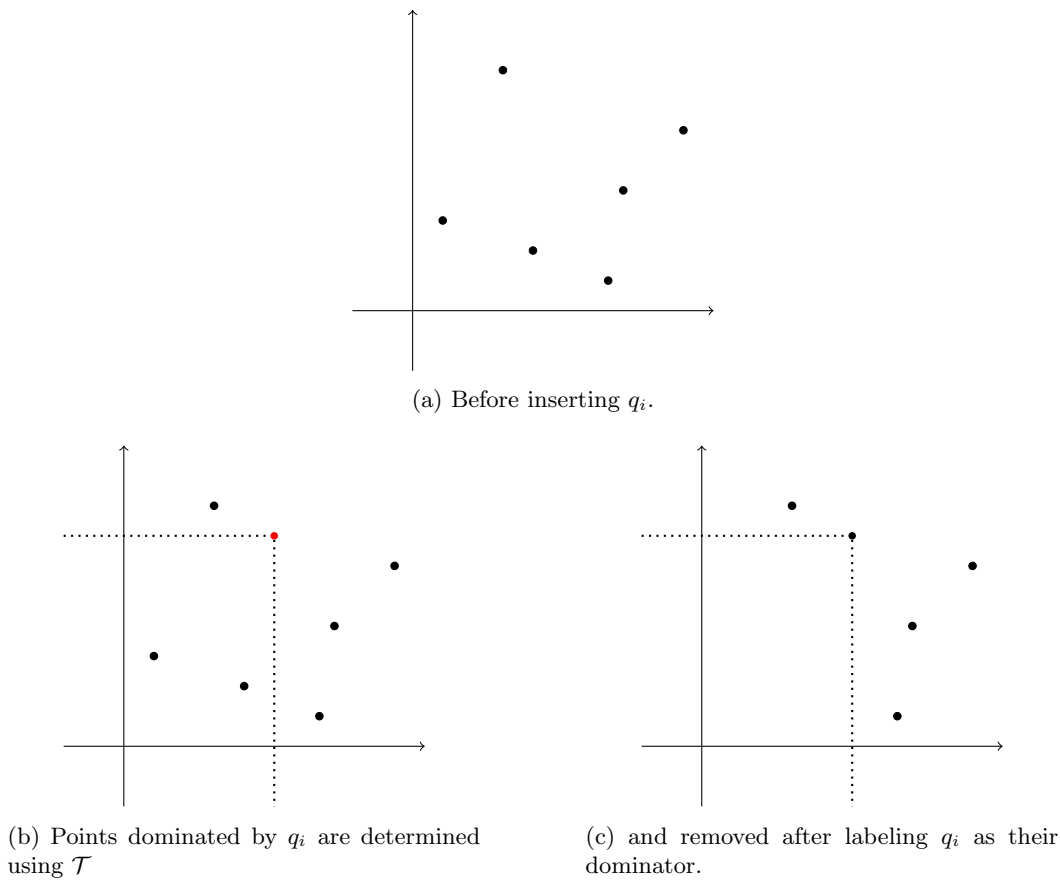
\begin{figure}
    \centering
    \begin{subfigure}{.45\textwidth}
        \centering
        \begin{tikzpicture}[scale=.4]
            \draw[->] (-2, 0) -- (10, 0);
            \draw[->] (0, -2) -- (0, 10);
            \foreach \x/\y in {1/3, 3/8, 4/2, 6.5/1, 7/4, 9/6} {
                \node[draw,circle,inner sep=1pt,fill]
                    at (\x, \y) {};
            }
        \end{tikzpicture}
        \caption{Before inserting $q_i$.}
    \end{subfigure}%
    \\[1em]
    \begin{subfigure}{.45\textwidth}
        \centering
        \begin{tikzpicture}[scale=.4]
            \draw[->] (-2, 0) -- (10, 0);
            \draw[->] (0, -2) -- (0, 10);
            \foreach \x/\y in {1/3, 3/8, 4/2, 6.5/1, 7/4, 9/6} {
                \node[draw,circle,inner sep=1pt,fill]
                    at (\x, \y) {};
            }
            \draw[dotted, thick] (-2, 7) -- (5, 7) -- (5, -2);
            \node[circle,inner sep=1pt,fill=red] at (5, 7) {};
        \end{tikzpicture}
        \caption{Points dominated by $q_i$ are determined using $\mcT$}
    \end{subfigure}%
    \hfill%
    \begin{subfigure}{.45\textwidth}
        \centering
        \begin{tikzpicture}[scale=.4]
            \draw[->] (-2, 0) -- (10, 0);
            \draw[->] (0, -2) -- (0, 10);
            \foreach \x/\y in {3/8, 6.5/1, 7/4, 9/6} {
                \node[draw,circle,inner sep=1pt,fill]
                    at (\x, \y) {};
            }
            \draw[dotted, thick] (-2, 7) -- (5, 7) -- (5, -2);
            \node[circle,inner sep=1pt,fill] at (5, 7) {};
        \end{tikzpicture}
        \caption{and removed after labeling $q_i$ as their dominator.}
    \end{subfigure}
    \caption{The algorithm for \cref{prb:midterm} in action.}
    \label{fig:midterm}
\end{figure}

\begin{problem} \label{prb:midterm}
    Let $P = \set{p_1, p_2, \dots, p_n} \subseteq \R$ be a set of $n$ points.
    Call this the \emph{real space}.
    Let $Q = \set{q_1, q_2, \dots, q_n} \subseteq \R^2$ be the \emph{feature space}.
    We call $q_i$ the feature point corresponding to $p_i$.
    For points $p_i$ and $p_j$, we say that $p_i$ \emph{dominates} $p_j$ iff
    $q_i \succ q_j$ in the feature space.
    For each $1 \le i \le n$, determine the closest point to $p_i$ (from $P$)
    that dominates it, or report that no such point exists.
\end{problem}
An $O(n^2)$ solution is immediate.
For each $p \in P$, iterate over all other points, and maintain the minimum distance among all
points that dominate it.
We give an $O(n \log n)$ solution.
Searching for better is futile, since sorting can be reduced to this problem.%
\footnote{Given an unsorted list $X = [x_1, x_2, \dots, x_n]$,
solve the given problem with $p_i = x_i$ and $q_i = (x_i, x_i)$.
Locate the minimum $x_i \in X$ and follow the nearest dominator to build a sorted list.}
\begin{solution}
    Given a point $p_i$, it suffices to solve for the closest point \emph{to the right of} $p_i$
    that dominates it, and the closest such point to the left.
    The closest point that dominates it will be one of these, which is easy to check.
    We give an algorithm to find the closest dominator to the right.
    Once the closest dominator on either side has been located, simply check which is closer to $p_i$.
    
    Relabel the points so that $p_1 < p_2 < \dots < p_n$.
    Maintain a balanced priority search tree $\mcT$, initially empty.
    Insert points $q_1, q_2, \dots$ one by one, and when inserting $q_i = (x, y)$, search the range
    $(-\infty, x] \times (-\infty, y]$ in $\mcT$.
    For any point $q_j$ in this range, mark $p_i$ as the closest dominator (to the right) of $p_j$
    and delete $q_j$ from $\mcT$.
    See \cref{fig:midterm} for an illustration of this step.

    \cite{DBLP:journals/siamcomp/McCreight85}
    shows how to maintain balanced priority search trees.
    Querying the range using $\mcT$ takes $O(\log n + k)$ time,
    where $k$ is the number of points in the range.
    Deleting each of these points takes a further $O(\log n)$ time per point, and inserting
    $q_i$ takes another $O(\log n)$.
    Since a point once deleted is never re-inserted, we get that the overall time complexity
    is $O(n \log n)$.
\end{solution}

We seek to generalize this to higher dimensions.
\begin{problem} \label{prb:general}
Consider the same problem as before but with $P \subseteq \R^2$
and $Q \subseteq \R^d$.
\end{problem}

Our earlier approach of splitting the problem into two directions no longer works.
In one dimension, we could write $\abs{p_i - p_j} = \max(p_i - p_j, p_j - p_i)$.
No such decomposition is available to us in higher dimensions.
Thus we turn to Voronoi diagrams.

\begin{solution}
    Construct a $d$-dimensional range tree $\mcT$ on the feature space $Q$.
    We view this as being sorted by the first coordinate in the first level,
    by the second coordinate in the second level, and so on.
    For each last-level subtree $t$ of $\mcT$, storing points
    $q_{i_1}, q_{i_2}, \dots$, pre-compute a Voronoi diagram on the sites
    $p_{i_1}, p_{i_2}, \dots$ and store it at the root of $t$.
    That is, store a Voronoi diagram on the real points corresponding to each
    feature point stored in the subtree.
    Construct a point location data structure from these Voronoi diagrams
    (e.g. trapezoidal maps) that allows querying in $O(\log n)$ time.
    This allows us to perform the following query:
    
    \begin{center}\begin{minipage}{.8\textwidth}
        Given a point $p \in \R^2$ and a rectangle $R \subseteq \R^d$,
        report the point $p_j$ in $P$ such that
        \begin{enumerate*}[(a)]
            \item $q_j \in R$, and
            \item $p_j$ is the closest to $p$ among all such points.
        \end{enumerate*}
    \end{minipage}\end{center}
    
    This is done by searching for this range using the range tree $\mcT$,
    but instead of reporting each point from the last-level canonical subtrees,
    query the point location structure stored at their roots.
    This gives us one candidate from each of the $O(\log^d n)$ canonical subtrees,
    and we report the one closest to $p$.

    For each $p_i \in P$, we are interested in the quadrant to the top and
    right of $q_i = (q_i^{(1)}, q_i^{(2)}, \dots, q_i^{(d)})$ (the rectangle
    $(q_i^{(1)}, \infty) \times (q_i^{(2)}, \infty) \times \dots \times (q_i^{(d)}, \infty)$).
    Perform this query for each $p_i$.

    The range tree can be constructed in $n \log^d n$ time.
    The Voronoi diagram/trapezoidal map on $k$ sites take $O(k \log k)$ time
    to construct.
    If $k_t$ is the number of points stored in the subtree $t$, then the total
    time for construction is of the order of \[
        \sum_t k_t \log(k_t) \le \log n \sum_t k_t = O(n \log^{d+1} n)
    \] where the sum ranges over all last-level subtrees of $\mcT$.
    $\sum_t k_t = O(n \log^d n)$ since each point appears in $O(\log^d n)$ last-level
    subtrees.

    Each query involves querying the nearest-neighbor in $O(\log^d n)$ canonical
    last-level subtrees.
    Each such query requires $O(\log n)$ time.
    Thus querying for all $p_i$ requires $O(n \log^{d+1} n)$ time.
\end{solution}

We have a couple of remarks to make regarding this algorithm.
First, we are in reality solving the more general:
\begin{problem} \label{prb:query}
    Let $P = \set{p_1, p_2, \dots, p_n} \subseteq \R^2$ be the real space and
    Let $Q = \set{q_1, q_2, \dots, q_n} \subseteq \R^d$ be the feature space.
    For any point $p \in \R^2$ and a rectangle $R \subseteq \R^d$, report the
    pair $(p_i, q_i)$ such that $q_i \in R$ and $p_i$ is the closest to $p$
    among all such pairs.
\end{problem}
All the algorithm does is use this as a black box for $n$ different queries.
This is more general in two senses:
\begin{itemize}
    \item The query point $p$ need not be from $P$.
    \item The rectangle $R$ need not be unbounded on any side.
\end{itemize}
In keeping with the restaurant setting mentioned in the introduction,
this would mean that \emph{any} user can query for the nearest restaurant
fitting their requirements.
For example, restaurants rated high on ambience and in some specified range of
affordability.

Secondly, point location in higher than $2$ dimensions requires super-exponential
time with respect to the dimension.
The naive algorithm already works in $O(n^2)$ time ($O(n^2 d')$ if the real space
has dimension $d'$---comparing two points and computing distances takes $O(d')$ time),
so generalizing this algorithm to higher dimensions of the real space offers no benefits.

\section{Can we make this faster?}
The presented algorithm has two phases---the construction of the range tree and point
location structures, and querying the constructed structure for each $p_i$.
\textit{Both} phases take $O(n \log^{d+1} n)$ time.
Any optimization must target both the phases.
However, optimizing query times is useful in regards to \cref{prb:query} even if
the construction time is not improved (or even compromised). \\

Our first idea was aimed at optimizing constructing the Voronoi diagrams.
Let $\mcV(v)$ denote the Voronoi diagram stored at any node $v$.
If we have computed the Voronoi diagram for each child of a node $v \in \mcT$,
say $v_1$ and $v_2$, can we merge them together to compute $\mcV(v)$?
Simply inserting all points from $\mcV(v_1)$ into $\mcV(v_2)$ incrementally
will not suffice, as this will still consume $O(n \log n)$ expected time
using the randomized incremental algorithm for Delaunay triangulations.
The merging step needs to be more efficient.

However, the range tree $\mcT$ is built on the feature space, and the Voronoi diagrams on the real space.
Though points in the feature space may be cleanly divided among the children by
$x$- or $y$-coordinate, the same points in the feature space could be woefully
intermingled.
We did not think this to be a promising line of attack, and did not pursue it further. \\

We tried to design an approximate algorithm, and understood that obtaining a multiplicative approximation factor is infeasible---for instance, when a query point is infinitesimally close to its closest point in $P$.

\subsection{Dynamic Nearest Neighbor Structures}
We remarked how in solving \cref{prb:general}, we have used a solution to \cref{prb:query},
which allows for much more general queries than we need.

We will now present an algorithm with a more offline flavor, that forgoes both generalities
of \cref*{prb:query},
capitalizing on the fact that we know all query points beforehand
and that our queries are always unbounded.
We will describe this for $d = 2$ for ease of presentation, and it extends readily to higher dimensions.

The idea is to sort and relabel the points so that $y_{q_1} > y_{q_2} > \dots > y_{q_n}$
and build only a $1$-dimensional range tree $\mcT_x$ on $Q$ based on the $x$-coordinate.
We maintain a \emph{dynamic} nearest-neighbor data structure $\NN(t)$ in each subtree $t$
of $\mcT_x$, initially empty.
We maintain the invariant that when we visit $p_i$ (in the specified order),
the nearest-neighbor structures contain precisely the $p_j$s such that
$q_j$ dominates $q_i$ in the $y$-coordinate.
Thus, a query $(x_{q_i}, \infty) \times (y_{q_i}, \infty)$ reduces to just
$(x_{q_i}, \infty)$ in $\mcT_x$.
From each canonical node $t$ of $\mcT_x$ for this query,
$\NN(t)$ is queried to collect one candidate nearest-neighbor,
and the closest neighbor is computed among the $O(\log n)$ candidates.
We then insert $p_i$ into $\NN(t)$ for each subtree $t$ containing $q_i$.

\begin{algo}
    \Fn{NearestDominator}{$P$, $Q$}
        \State sort $Q$ by $y$-coordinate
        \State build the range tree $\mcT_x$
        \State for each subtree $t$, initialize $\NN(t) \gets \varnothing$
        \For{$i \gets 1$ to $n$}
            \State query the range $[x_{q_i}, \infty)$ in $\mcT_x$ to find the nearest dominator of $p_i$
            \For{each $t$ containing containing $q_i$}
                \State insert $q_i$ into $\NN(t)$
            \EndFor
        \EndFor
    \EndFn
\end{algo}

This removes one dimension from the range trees, but adds complexity to
the nearest-neighbor structure in order to support dynamic insertions.
There are only $O(n)$ nodes in $\mcT_x$, and only $O(\log n)$ canonical subtrees for any query.
If the dynamic nearest-neighbor structure $\NN$ supports $O(\log n)$ insertion and
$O(\log n)$ query time, we would have reduced the overall time complexity to
$O(n \log^2 n)$ (or in the general case where $Q \subseteq \R^d$, $O(n \log^d n)$).
If the distributions of $P$ and $Q$ are largely independent of each other, sorting by $y$-coordinate
in the feature space would lead to a nearly uniform random ordering of $P$.

\textit{On choosing the Nearest Neighbors data structure.}
The literature suggests that KD-trees and R*-trees support $O(\log n)$
nearest-neighbor queries in expectation when points are randomly distributed 
\cite{10.1145/355744.355745}.
It is not fully clear to us how well these support dynamic insertions.

We looked at \cite{10.1145/1706591.1706596} for a dynamic
Voronoi diagram construction, but the time bounds of $O(\log^3 n)$ for insertion and $O(\log^2 n)$
for deletion are both too large to be helpful.

\bibliographystyle{alpha}
\bibliography{citations}

\end{document}